\documentclass[namedreferences]{spackap}

\usepackage{graphicx} 
\usepackage{klups} 

\begin{document}

\begin{article}

\begin{opening}

\title{Is planetary migration inevitable?}
\runningtitle{Is planetary migration inevitable?}

\author{Caroline
E.~J.~M.~L.~J. \surname{Terquem}} 
\runningauthor{C. Terquem} 
\institute{Institut d'Astrophysique de Paris, 98 bis, bd Arago,
F--75014 Paris \\ Universit\'e Denis Diderot--Paris VII, 2 Place
Jussieu, F--75251 Paris Cedex 5}

\received{\ldots} \accepted{\ldots}

\begin{ao}
C. Terquem, Institut d'Astrophysique de Paris, 98 bis, bd Arago,
F--75014 Paris, terquem@iap.fr
\end{ao}

\begin{abstract}
According to current theories, tidal interactions between a disk and
an embedded planet may lead to the rapid migration of the protoplanet
on a timescale shorter than the disk lifetime or estimated planetary
formation timescales.  Therefore, planets can form only if there is a
mechanism to hold at least some of the cores back on their way in.
Once a giant planet has assembled, there also has to be a mechanism to
prevent it from migrating down to the disk center.  This paper reviews
the different mechanisms that have been proposed to stop or slow down
migration.
\end{abstract}

\end{opening}

\section{Introduction}

Almost 20\% of the extrasolar planets detected so far orbit at a
distance between 0.038 and 0.1 astronomical unit (au) from their host
star.  

It is very unlikely that these short--period giant planets, also
called 'hot Jupiters', have formed {\it in situ}.  In most of the
standard disk models, temperatures at around 0.05~au are larger than
1500~K (Bell et al. 1995; Papaloizou \& Terquem 1999), preventing the
condensation of rocky material and therefore the accretion of a solid
core there.  Even if models with lower temperatures are considered,
giant planets may form in close orbits according to the core accretion
model only if the disk surface density is rather large and the
accretion process very efficient (Bodenheimer et al. 2000; Ikoma et
al. 2001).  This is because at 0.05~au a solid core of about 40 earth
masses has to be assembled before a massive gaseous envelope can be
accreted (Papaloizou \& Terquem 1999; Bodenheimer et al. 2000).  More
likely, the hot Jupiters have formed further away in the
protoplanetary nebula and have migrated down to small orbital
distances.  It is also possible that migration and formation were
concurrent (Papaloizou \& Terquem 1999).

So far, three mechanisms have been proposed to explain the location of
planets at very short orbital distances.  One of them relies on the
gravitational interaction between two giant planets, which may lead to
orbit crossing and to the ejection of one planet while the other is
left in a smaller orbit (Rasio \& Ford 1996; Weidenschilling \&
Marzari 1996).  However, this mechanism cannot reproduce the orbital
characteristics of the extrasolar planets observed so far.  Another
mechanism is the so--called 'migration instability' (Murray et
al. 1998; Malhotra 1993).  It involves resonant interactions between
the planet and planetesimals located inside its orbit which lead to
the ejection of a fraction of them while simultaneously causing the
planet to migrate inward.  Such interactions require a very massive
disk to move a Jupiter mass planet from several astronomical units
down to very small radii, as there has to be at least on the order of
a Jupiter mass of planetesimals inside the orbit of the planet.  Such
a massive disk is unlikely and furthermore it would be only marginally
gravitationally stable.  The third and most efficient mechanism
involves the tidal interaction between the protoplanet and the gas in
the surrounding protoplanetary nebula (Goldreich \& Tremaine 1979,
1980; Lin \& Papaloizou 1979, 1993 and references therein; Papaloizou
\& Lin 1984; Ward 1986, 1997a).  Here again the protoplanet can move
significantly only if there is at least a comparable mass of gas
within a radius comparable to that of its orbit. However this is not a
problem since this amount of gas is needed anyway in the first place
to form the planet.

Note that hot Jupiters can also be produced by the dynamical
relaxation of a population of planets on inclined orbits, formed
through gravitational instabilities of a circumstellar envelope or a
thick disk (Papaloizou \& Terquem 2001).  However, if objects as
heavy as $\tau$--Boo may be produced {\it via} fragmentation, it is
unlikely that lower mass objects would form that way.

Tidal interaction between a disk and a planet may lead to two
different types of migration (Ward 1997a; Terquem et al. 2000 and
references therein).  Cores with masses up to about 10~M$_\oplus$
interact linearly with the surrounding nebula and migrate inward {\em
relative} to the gas (type~I migration).  Planets with masses at least
comparable to that of Jupiter interact nonlinearly with the disk and
may open up a gap (Goldreich \& Tremaine 1980; Lin \& Papaloizou 1979,
1993 and references therein).  The planet is then locked into the
angular momentum transport process of the disk, and migrates {\em
with} the gas at a rate controlled by the disk viscous timescale
(type~II migration).  The direction of type~II migration is that of
the viscous diffusion of the disk.  Therefore it is inward except in
the outer parts of the disk which diffuse outward.

The drift timescale for a planet of mass $M_{\rm{pl}}$ undergoing
type~I migration in a uniform disk is (Ward 1986, 1997a):

\begin{equation}
\tau_{\rm{I}} (\rm{yr}) \sim 10^8 
\left( \frac{M_{pl}}{\rm{M}_{\oplus}} \right)^{-1} 
\left( \frac{\Sigma}{\rm{g~cm}^{-2}} \right)^{-1}
\left( \frac{r}{\rm{au}} \right)^{-1/2}
\times 10^2 \left(\frac{H}{r} \right)^2
\label{tauI}
\end{equation}

\noindent where $\Sigma$ is the disk surface density, $r$ is the
distance to the central star, and $H$ is the disk semithickness.  It
is assumed here that the torque exerted by the material which
corotates with the perturbation can be neglected.  

For type~II migration, the characteristic orbital decay timescale is
the disk viscous timescale:

\begin{equation}
\tau_{\rm{II}} (\rm{yr}) = \frac{1}{ 3\alpha} \left( \frac{r}{H} \right)^2
\Omega^{-1} = 0.05 \frac{1}{\alpha} \left( \frac{r}{H} \right)^2 \left(
  \frac{r}{\rm{AU}} \right)^{3/2} 
\label{tauII}
\end{equation}
 
\noindent where $\alpha$ is the standard Shakura \& Sunyaev (1973)'s
parameter and $\Omega$ is the angular velocity at radius $r$.

It has recently been shown by Masset \& Papaloizou (2003) that a
planet in the intermediate mass range embedded in a disk massive
enough may undergo a runaway migration.  This typically happens for
Saturn--sized giant planets embedded in disks with a mass several
times the minimum mass of the solar nebula.  The timescale for runaway
migration can be much shorter than that for type~I or type~II
migration.

For typical disk parameters, the timescales given above are much
shorter than the disk lifetime or estimated planetary formation
timescales.  Therefore, planets can form only if there is a mechanism
to hold at least some of the cores back on their way in.  Once a giant
planet has assembled, there also has to be a mechanism to prevent it
from migrating down to the disk center.  We now review the different
mechanisms that have been proposed to slow down, stop or reverse
migration.

\section{Stopping type~I migration}

\subsection{Stopping type~I migration at small radii}

Cores undergoing orbital decay due to type~I migration would stop
before plunging onto the star if the disk had an inner
(magnetospheric) cavity.  Merger of incoming cores and subsequent
accretion of a massive gaseous atmosphere could then produce a hot
Jupiter {\em in situ} (Ward 1997b, Papaloizou \& Terquem 1999).
However, the extent of magnetospheric cavities is very limited, and
therefore this mechanism cannot account for the presence of planets
orbiting further away from the central star.  For these planets to be
assembled, type~I migration has to be either avoided or halted.
Type~I migration would not take place if the interaction between the
core and the disk could become nonlinear with a resulting gap
formation.  However, as discussed by Terquem et al.  (2000), this
situation is unlikely.  We now review the mechanisms that have been
proposed to halt or reverse type~I migration.

\subsection{Migration of planets on eccentric orbits}

The migration timescale given by equation~(\ref{tauI}) applies to a
planet on a circular orbit.  Papaloizou \& Larwood (2000) have
investigated the case of a planetary core on an eccentric orbit (in an
axisymetric disk) with an eccentricity $e$ significantly larger than
the disk aspect ratio $H/r$.  They found that the direction of orbital
migration reverses for fixed eccentricity $e>1.1 H/r$.  This is
because the core spends more time near apocenter, where it is
accelerated by the surrounding gas, than near pericenter, where it is
decelerated.  In general, the interaction between the core and the
disk leads to eccentricity damping (Goldreich \& Tremaine 1980).
However, Papaloizou \& Larwood (2000) showed that a significant
eccentricity could be maintained by gravitational interactions with
other cores.  Papaloizou (2002) further studied the case of a core
embedded in an eccentric disk.  He showed that migration may be
significantly reduced or even reverse from inward to outward when the
eccentricity of the orbit of the core significantly exceeds that of
the disk when that is large compared to $H/r$ and the density profile
is favorable.  In some cases, such a high orbital eccentricity may be
an equilibrium solution, and therefore suffers no damping.

\subsection{Stopping migration by a toroidal magnetic field }

Terquem~(2003) has investigated the effect of a toroidal magnetic
field on type~I migration for a planet on a circular orbit.  When a
field is present, in contrast to the nonmagnetic case, there is no
singularity at the corotation radius, where the frequency of the
perturbation matches the orbital frequency.  However, all fluid
perturbations are singular at the so--called {\em magnetic
resonances}, where the Doppler shifted frequency of the perturbation
matches that of a slow MHD wave propagating along the field line.
There are two such resonances, located on each side of the planet's
orbit and within the Lindblad resonances. Like in the nonmagnetic
case, waves propagate outside the Lindblad resonances.  But they also
propagate in a restricted region around the magnetic resonances.

The magnetic resonances contribute to a significant global torque
which, like the Lindblad torque, is negative (positive) inside
(outside) the planet's orbit.  Since these resonances are closer to
the planet than the Lindblad resonances, they couple more strongly to
the tidal potential and the torque they contribute dominates over the
Lindblad torque if the magnetic field is large enough.  In addition,
if $\beta \equiv c^2/v_A^2$, where $c$ is the sound speed and $v_A$
the Alfven velocity, increases fast enough with radius, the outer
magnetic resonance becomes less important (it disappears altogether
when there is no magnetic field outside the planet's orbit) and the
total torque is then negative, dominated by the inner magnetic
resonance.  This leads to outward migration of the planet.



The amount by which $\beta$ has to increase outward for the total
torque exerted on the disk to be negative depends mainly on the
magnitude of $\beta$.  It was found that, for $\beta =1$ or 100 at
corotation, the torque exerted on the disk is negative when $\beta$
increases at least as fast as $r^2$ or $r^4$, respectively.



The migration timescales that correspond to the torques calculated
when a magnetic field is present are rather short.  The orbital decay
timescale of a planet of mass $M_{\rm{pl}}$ at radius $r_p$ is $\tau =
M_p r_p^2 \Omega_p / |T|$, where $T$ is the torque exerted by the
planet on the disk and $\Omega_p$ is the angular velocity at radius
$r_p$.  This gives:

\begin{equation}
\tau ({\rm yr}) = 4.3 \times 10^9 \left( \frac{M_{\rm{pl}}}{{\rm
M_\oplus}} \right)^{-1} \left( \frac{ \Sigma_p }{100 \; {\rm g \;
cm^{-2}}} \right)^{-1} \left( \frac{r_p}{1 \; {\rm au}} \right)^{-1/2}
\left( \frac{|T|}{\Sigma_p r_p^4 \Omega_p^2} \right)^{-1}
\end{equation}

\noindent where $\Sigma_p$ is the disk surface density at radius
$r_p$.  In a standard disk model, $\Sigma \sim
100$--$10^3$~g~cm$^{-2}$ at 1~au (see, for instance, Papaloizou \&
Terquem~1999).  Therefore, $\tau \sim 10^5$--$10^6$~yr for a one earth
mass planet at 1~au in a nonmagnetic disk, as $|T| / \left(\Sigma_p
r_p^4 \Omega_p^2 \right) \sim 10^3$ in that case (see fig.~9 from
Terquem~2003).  This is in agreement with Ward (1986, 1997a, see
eq.~[\ref{tauI}] above).  In a magnetic disk, $|T| / \left(\Sigma_p
r_p^4 \Omega_p^2 \right)$ may become larger (see fig.~9 from
Terquem~2003) leading to an even shorter migration timescale.
However, it is important to keep in mind that these timescales are
{\em local}.  Once the planet migrates outward out of the region where
$\beta$ increases with radius, it may enter a region where $\beta$
behaves differently and then resume inward migration for instance.

The calculations summarized here indicate that a planet migrating
inward throu\-gh a nonmagnetized region of a disk would stall when
reaching a magnetized region.  It would then be able to grow to become
a terrestrial planet or the core of a giant planet.  We are also led
to speculate that in a turbulent magnetized disk in which the large
scale field structure changes sufficiently slowly a planet may
alternate between inward and outward migration, depending on the
gradients of the field encountered.  Its migration could then become
diffusive, or be limited only to small scales.

\section{Stopping type~II migration}

\subsection{Stopping type~II migration at small radii}

Like cores undergoing type~I migration, planets subject to type~II
migration would stop before falling onto the star if they entered a
magnetospheric cavity.  Tidal interaction with a rapidly rotating star
would also halts planet orbital decay at a few stellar radii (where
the interaction becomes significant; see Lin et al. 2000 and
references therein).  Both of these mechanisms have been put forth to
account for the present location of the planet around 51~Pegasi (Lin
et al. 1996), and to explain the location of hot Jupiters more
generally.

A planet overflowing its Roche lobe and losing part of its mass to the
central star would also halt at small radii.  This is because during
the transfer of mass the planet moves outward to conserve the angular
momentum of the system (Trilling et al. 1998).  The planet stops at
the location where its physical radius is equal to its Roche radius.
Recent observations of atomic hydrogen absorption in the stellar
Lyman~$\alpha$ line during three transits of the planet HD209458b
suggest that hydrogen atoms are escaping the planetary atmosphere
(Vidal--Madjar et al. 2003).

\subsection{``The last of the Mohicans''...}

The mechanisms reviewed above cannot account for the presence of giant
planets orbiting their parent star at distances larger than about a
tenth of an au.

It has been suggested that migration of a giant planet could be
stopped at any radius if migration and disk dissipation were
concurrent (Trilling et al. 1998, 2002).  Note that a massive planet
can suffer significant orbital decay only if the mass of gas in its
vicinity is comparable to the mass of the planet itself.  If the disk
is significantly less massive, there is not enough gas around the
planet to absorb its angular momentum, and migration is slown down
(Ivanov et al. 1999).  If the disk dissipates while migration is
taking place, then the drift timescale may increase in such a way that
the planet stalls at some finite radius.  Note however that this
requires very fine tuning of the parameters (disk mass, disk lifetime
etc.), as for a given disk mass the migration timescale decreases as
the orbital radius decreases (see eq.[\ref{tauII}] above).  Also, a
major problem with this mechanism is to explain how the disk
dissipates.  Within this scenario, there is initially enough gas in
the disk to push the planet down to some orbital radius.  For typical
disk parameters, only part of this gas may be accreted by the planet
or leak through the gap to be accreted onto the star (Bryden et
al. 1999, Kley 1999).  It is therefore not clear how the gas
disappears.

A giant planet could survive if after it formed there were not enough
material left in the disk for significant migration to occur.  It has
been suggested that a series of giant planets could actually assemble
in the disk and disappear onto the star (Gonzalez 1997, Laughlin \&
Adams 1997).  Then at some point the disk mass may be such that one
more planet can be formed but not migrate (Lin 1997).  This survivor
is sometimes refereed to as the last of the Mohicans.

\subsection{Planets locked in resonances}

Within the context of several planets forming in a disk, another
scenario has been suggested to occur when the migration of the
innermost planet is stopped by either the star's tidal barrier or a
magnetospheric cavity.  In that case, a second planet approaching the
star would stop when entering a low order resonance with the innermost
planet, i.e. when the mean motions of the two planets become
commensurate.  As shown by Goldreich (1965) in the context of our
Solar system, such commensurabilities are stable because angular
momentum is secularly transferred between the different objects in
just the correct proportion to keep the mean motions commensurate.  In
the context of the planetary system discussed here, the angular
momentum would be transferred from the central star to the innermost
planet then to the ring of material trapped between this planet and
the next one, then from this ring to the next planet, and so on until
the angular momentum is transferred to the disk outer part.  The
evolution of the central star may cause the whole system to migrate
either inward or outward, but the planets remain locked into the
resonances (Lin 1997).

Like with the scenario discussed in the previous subsection, a major
problem here is to explain how the disk material trapped in between
the different planets eventually disappears.

Resonant trapping of planets can also lead to a reversal of type~II
migration.  This happens when a giant planet (e.g. Jupiter) migrating
inward captures into the 2:3 resonance a lighter outer giant planet
(e.g. Saturn, Masset \& Snellgrove 2001).  The gaps that the two
planets open in the disk then overlap, and the imbalance between the
torque exerted at Jupiter's inner Lindblad resonance and that exerted
at Saturn's outer Lindblad resonance causes the whole system to
migrate outward.  This outward migration is accompanied by an
increased mass flow through the overlapping gaps, as the angular
momentum gained by the planets is lost by the disk.

\section{Conclusions}

We have reviewed the different mechanisms which have been proposed to
slow down, halt or reverse inward orbital migration.  When the
interaction between the disk and the planet(s) is linear, orbital
decay can be stopped if a magnetic field is present in the disk or if
the planets are on sufficiently eccentric orbits.  For a nonlinear
interaction, no general mechanism has been shown to prevent orbital
decay of an isolated planet.  Whether type~II migration occurs or not
may just depend on the mass of gas left in the disk after the planet
forms.

\end{article}


\begin{thebibliography}{}

\bibitem[]{} Bell, K. R., D.~N.~C.~Lin, L.~W.~Hartmann and
  S.~J.~Kenyon: 1995. 'The FU Orionis outburst as a thermal accretion
  event: Observational constraints for protostellar disk models', {\em
  Astroph. J.} {\bf 444}, 376--395

\bibitem[]{} Bodenheimer,P., O. Hubickyj and J.~J.~Lissauer: 2000,
  'Models of the in situ formation of detected extrasolar giant
  planets', {\em Icarus} {\bf 143} 2--14

\bibitem[]{} Bryden, G., X. Chen, D.~N.~C. Lin, R.~P. Nelson and
    J.~C.~B. Papaloizou: 1999, 'Tidally induced gap formation in
    protostellar disks: gap clearing and suppression of protoplanetary
    growth', {\em Astroph. J.} {\bf 514}, 344--367

\bibitem[]{} Goldreich, P.: 1965, 'An explanation of the frequent
  occurence of commensurable mean motions in the solar system', {\em
  M.N.R.A.S.} {\bf 130}, 159--181

\bibitem[]{} Goldreich, P. and S. Tremaine: 1979, 'The excitation of
    density waves at the Lindblad and corotation resonances by an external
    potential', {\em Astroph. J.} {\bf 233}, 857--871

\bibitem[]{} Goldreich, P. and S. Tremaine: 1980, 'Disk--satellite
    interactions', {\em Astroph. J.} {\bf 241}, 425--441

\bibitem[]{} Gonzalez, G.: 1997, 'The stellar metallicity--giant
  planet connection', {\em M.N.R.A.S.} {\bf 285}, 403--412

\bibitem[]{} Ikoma, M., H. Emori and K. Nakazawa: 2001, 'Formation of
  giant planets in dense nebulae: critical core mass revisited', {\em
  Astroph. J.} {\bf 553}, 999--1005

\bibitem[]{} Ivanov, P.~B., J.~C.~B. Papaloizou and A.~G. Polnarev:
    1999, 'The evolution of a supermassive binary caused by an
    accretion disc', {\em M.N.R.A.S.} {\bf 307}, 791--891

\bibitem[]{} Kley, W.: 1999, 'Mass flow and accretion through gaps in
    accretion disks', {\em M.N.R.A.S.} {\bf 303}, 696--710

\bibitem[]{} Laughlin, G. and F.~C. Adams: 1997, 'Possible stellar
  metallicity enhancements from the accretion of planets', {\em
  Astrophys. J.} {\bf 491}, L51--L54

\bibitem[]{} Lin, D.~N.C.: 1997, 'Planetary formation in protostellar
  disks', In {\it Accretion phenomena and related outflows \/},
  eds. D.~T. Wickramasinghe, G.~V. Bicknell, \& L. Ferrario (ASP
  Conf. Series, vol. 121), p.~321--330

\bibitem[]{} Lin, D.~N.~C., P. Bodenheimer and D.~C.~Richardson: 1996,
  'Orbital migration of the planetary companion of 51 Pegasi to its
  present location', {\it Nature} {\bf 380}, 606--607

\bibitem[]{} Lin, D.~N.~C. and J.~C.~B.~Papaloizou: 1979, 'On the
    structure of circumbinary accretion disks and the tidal evolution
    of commensurable satellites', {\em M.N.R.A.S.} {\bf 188},
    191--201

\bibitem[]{} Lin, D.~N.~C. and J.~C.~B.~Papaloizou: 1993, 'On the
  tidal interaction between protostellar disks and companions', In
  {\it Protostars and Planets III \/}, eds. E.~H. Levy, \&
  J.~I. Lunine (Tucson: University of Arizona Press), p. 749--835

\bibitem[]{} Lin, D.~N.~C., J.~C.~B.~Papaloizou, C. Terquem, G. Bryden
  and S. Ida: 2000, 'Orbital evolution and planet--star tidal
  interaction', In {\em Protostars and Planets IV \/},
  eds. V. Mannings, A.~P. Boss, \& S.~S. Russell (Tucson: University
  of Arizona Press), p.~1111

\bibitem[]{} Malhotra, R.: 1993, 'The origin of Pluto's peculiar
  orbit', {\it Nature\/} {\bf 365}, 819

\bibitem[]{} Masset, F. and J.~C.~B. Papaloizou: 2003, 'Runaway
  migration and the formation of hot Jupiters', {\em Astrophys. J.}
  {\bf 588}, 494--508

\bibitem[]{} Masset, F. and M. Snellgrove: 2001, 'Reversing type~II
  migration: resonance trapping of a lighter giant protoplanet', {\em
  M.N.R.A.S.}  {\bf 320}, L55--L59

\bibitem[]{} Murray, N., Hansen, B., Holman, M. and Tremaine, S.:
  1998, 'Migrating Planets', {\it Science \/} {\bf 279}, 69--72

\bibitem[]{} Papaloizou, J.~C.~B.: 2002, 'Global $m=1$ modes and
  migration of protoplanetary cores in eccentric protoplanetary
  discs', {\em Astron. Astrophys.} {\bf 388}, 615--631

\bibitem[]{} Papaloizou, J.~C.~B. and J.~D. Larwood: 2000, 'On the
  orbital evolution and growth of protoplanets embedded in a gaseous
  disc', {\em M.N.R.A.S.}  {\bf 315}, 823--833

\bibitem[]{} Papaloizou, J.~C.~B. and D.~N.~C.~Lin: 1984, 'On the
    tidal interaction between protoplanets and the primordial solar
    nebula.  I~--- Linear calculation of the role of angular momentum
    exchange', {\em Astroph. J.} {\bf 285}, 818--834

\bibitem[]{} Papaloizou, J. C. B. and C. Terquem: 1999, 'Critical
  protoplanetary core masses in protoplanetary disks and the formation
  of short--period giant planets', {\em Astroph. J.} {\bf 521},
  823--838

\bibitem[]{} Papaloizou, J. C. B. and C. Terquem: 2001, 'Dynamical
  relaxation and massive extrasolar planets', {\em M.N.R.A.S.}  {\bf
  325}, 221--230

\bibitem[]{} Rasio, F.~A., \&  E.~B. Ford: 1996, 'Dynamical
  instabilities and the formation of extrasolar planetary systems',
  {\it Science} {\bf 274}, 954--956.

\bibitem[]{} Shakura, N.~I. and R.~A.~Sunyaev: 1973, 'Black holes in
  binary systems: Observational appearance', {\em Astron. Astroph.}
  {\bf 24}, 337--355

\bibitem[]{} Terquem, C. E. J. M. L. J.: 2003, 'Stopping inward
  planetary migration by a toroidal magnetic field', {\em M.N.R.A.S.}
  {\bf 341}, 1157--1173

\bibitem[]{} Terquem, C., J.~C.~B. Papaloizou and R.~P. Nelson: 2000,
  ' Disks, extrasolar planets and migration', In {\it From Dust to
  Terrestrial Planets}, eds. W. Benz, R. Kallenbach, G. Lugmair \&
  F. Podosek (ISSI Space Sciences Series, 9, reprinted from Space
  Science Reviews, 92), p.~323--340

\bibitem[]{} Trilling, D.~E., W. Benz, T. Guillot, J.~I.~Lunine,
  W.~B.~Hubbard and A. Burrows: 1998, 'Orbital evolution and migration
  of giant planets: modeling extrasolar planets', {\em Astroph. J.}
  {\bf 500}, 428--439

\bibitem[]{} Trilling, D.~E., J.~I.~Lunine and W. Benz: 2002, 'Orbital
  migration and the frequency of giant planet formation', {\em
  Astron. Astroph.}  {\bf 394}, 241--251

\bibitem[]{} Vidal--Madjar, A., A. Lecavelier des Etangs,
  J.--M. D\'esert, G.~E. Ballester, R. Ferlet, G. H\'ebrard and
  M. Mayor: 2003, 'An extended upper atmosphere around the extrasolar
  planet HD209458b', {\em Nature} {\bf 422}, 143--146

\bibitem[]{} Ward, W.~R.: 1986, 'Density waves in the solar nebula --
    Differential Lindblad torque', {\em Icarus} {\bf 67}, 164--180

\bibitem[]{} Ward, W.~R.: 1997a, 'Protoplanet Migration by Nebula
    Tides', {\em Icarus} {\bf 126}, 261--281

\bibitem[]{} Ward, W.~R.: 1997b, 'Survival of Planetary Systems', {\em
    Astroph. J.} {\bf 482}, L211--L214

\bibitem[]{} Weidenschilling, S.~J., \& Marzari, F.: 1996,
  'Gravitational scattering as a possible origin for giant planets at
  small stellar distances', {\it Nature\/} {\bf 384}, 619--621



\end{thebibliography}
\end{document}